\documentclass[prl,aps,nofootinbib,superscriptaddress,showkeys,showpacs,twocolumn,10pt,floatfix]{revtex4-1}
\usepackage{epsfig}
\usepackage{graphicx}
\usepackage[small]{subfigure}
\usepackage{setspace}
\usepackage[T1]{fontenc}
\usepackage[utf8]{inputenc}
\usepackage[english]{babel}
\usepackage{mathrsfs}
\usepackage{braket}
\usepackage[overload]{empheq} 
\usepackage{amssymb}
\usepackage{amsmath}
\usepackage{amsfonts}
\usepackage{siunitx}
\usepackage{amsopn}
\usepackage{mathtools}
\usepackage{float}
\usepackage{chemformula}
\usepackage{comment} 
\usepackage{tikz}
\usepackage{booktabs}
\usepackage{multirow}
\usepackage{verbatim}
\usepackage{ulem}
\usepackage{soul}
\usetikzlibrary{arrows}
\usepackage[colorlinks,linkcolor=blue,urlcolor=blue,citecolor=blue]{hyperref}

\renewcommand{\sout}{\bgroup \color{red} \ULdepth=-.5ex \ULset}

\usepackage{tikz,xcolor,hyperref}
\definecolor{lime}{HTML}{A6CE39}
\DeclareRobustCommand{\orcidicon}{
	\begin{tikzpicture}
	\draw[lime, fill=lime] (0,0) 
	circle [radius=0.16] 
	node[white] {{\fontfamily{qag}\selectfont \tiny ID}};
	\draw[white, fill=white] (-0.0625,0.095) 
	circle [radius=0.007];
	\end{tikzpicture}
	\hspace{-2mm}
}

\foreach \x in {A, ..., Z}{%
	\expandafter\xdef\csname orcid\x\endcsname{\noexpand\href{https://orcid.org/\csname orcidauthor\x\endcsname}{\noexpand\orcidicon}}
}

\begin{document}

\title{Dynamics of dilute nuclear matter with light clusters and in-medium effects}

\author{Rui Wang\orcidA{}}
\email[]{rui.wang@ct.infn.it}
\affiliation{Istituto Nazionale di Fisica Nucleare~(INFN), Sezione di Catania, I-95123 Catania, Italy}
\author{Stefano Burrello\orcidB{}}
\email[]{burrello@lns.infn.it}
\affiliation{INFN, Laboratori Nazionali del Sud, I-95123 Catania, Italy}
\author{Maria Colonna\orcidC}
\email[]{colonna@lns.infn.it}
\affiliation{INFN, Laboratori Nazionali del Sud, I-95123 Catania, Italy}
\author{Francesco Matera}
\email[]{francesco.matera@unifi.it}
\affiliation{Dipartimento di Fisica e Astronomia, I-50019 Sesto Fiorentino, Firenze,
Italy}

\date{\today}

\begin{abstract}
\noindent 
We investigate the dynamics of dilute systems composed of nucleons and light clusters within a linear response approach, taking into account the in-medium Mott effects on cluster appearance, through a density-dependent momentum cut-off. We find that spinodal instabilities and associated growth rates are severely affected by the presence of light clusters and, in particular, by the treatment of in-medium effects, foreshadowing intriguing consequences for fragment formation in heavy-ion collisions and in the broader astrophysical context.
\end{abstract}

\maketitle

\paragraph{Introduction.}
The understanding of the composition and of the thermodynamical properties of nuclear matter, including the possible occurrence of phase transitions and condensates, is of crucial importance in several areas of nuclear physics, playing a considerable role also in astrophysics and cosmology~\cite{HorowitzNPA2006, typelPRC2010, GulminelliPRC2015, BausweinPRL2019, BurgioPPNP2021, SedrakianPPNP2023, ZhouNAT2023}. In particular, the properties of nuclear (and neutron) matter at very low densities have recently attracted a lot of interest~\cite{burrelloEPJA2022, kellerPRL2023}.
For instance, they have a considerable impact on the characteristics of exotic neutron-rich nuclei~\cite{TypelPRC2014, burrelloPRC2021}, as well as on the dynamics of supernova collapse, influencing  the possible emission of neutrino and gravitational wave signals~\cite{TewsEPJA2019, OertelPRC2020} and the structure of proto-neutron stars~\cite{burrelloPRC2015, ThiAA2021, gramsEPJA2022}. 

In the density regime below the saturation value ($\rho_{0}$), nucleon correlations are expected to play a leading role~\cite{riosPRC2014}. A typical example of the emergence of large-scale correlations are fragmentation processes, intimately connected to liquid-gas phase transitions and the occurrence of volume (spinodal) instabilities~\cite{ChomazPR2004}. Few-body correlations remain important even at lower densities ($\lesssim 10^{-2}\rho_{0}$), as the system can minimize its energy by forming light clusters such as deuterons, or strongly bound $\alpha$ particles~\cite{typelPRC2010}. Up to moderate temperatures, due to Pauli-blocking effects exerted on their constituent nucleons by  the surrounding nuclear medium, the effective binding energy of the clusters is expected to decrease with density until it vanishes and the clusters dissolve (the so-called Mott effect), no longer being bound~\cite{ropkePRC2015}. Hence, owing to the delicate interplay between mean-field instabilities, nucleon correlations and in-medium effects, a unified theoretical understanding of the composition of nuclear matter as a function of density and temperature still represents a true challenge. Thermodynamical approaches aiming at a fully consistent description of the concurrent appearance of light clusters and heavier fragments have recently been proposed~\cite{hempelASJ2012, paisPRC2019, RoepkePRC2020}.

On the other hand, a relevant source of information on the nuclear Equation of State comes from the study of heavy-ion collisions, which represent a unique tool to create, in terrestrial laboratories, transient states, possibly locally equilibrated, of nuclear matter under several conditions away from saturation
values~\cite{baranPREP2005, liPREP2008}. In particular, central collisions at Fermi/intermediate energies lead to compression in the initial stage, which allow to probe the high-density regime by experimental observables such as collective flows and particle/meson production, but also to explore low-density regions, in the subsequent expansion phase~\cite{ColonnaPPNP2020, SorensenPPNP2024}.  Being, in general, out of equilibrium processes, heavy-ion reactions are usually modeled with transport theories~\cite{TMEP, xuPPNP2019}; among them, only a few approaches include, beside nucleons, light-nuclei as explicit degrees of freedom~\cite{DanNPA533, OnoJPCS2013, OliPRC99, onoPPNP2019, WangPRC2023, CociPRC2023, chengPRC2024, SKJNC2024}. Specifically, the production and dissociation of the deuteron, triton, $^3$He, and $\alpha$ particles appear in the formalism of Ref.~\cite{WangPRC2023} as many-particle scatterings. On the other hand, stochastic approaches of the Boltzmann-Langevin type~\cite{ChomazPR2004, NapolitaniPLB2013, LinPRC2018, ColonnaPPNP2020, TMEP} have shown to well reproduce the emergence of intermediate mass fragments as resulting from the occurrence of spinodal instabilities. The formulation of transport theories accounting on equal footing and in a consistent manner for the description of few-body correlations and mean-field instabilities would be thus highly desirable. In such a context, in this Letter we present a novel approach to solve, within a linear response framework, the nuclear dynamics in the heterogeneous sub-saturation regime. In particular, we investigate the fragmentation dynamics of a system initialized at low density and at a given temperature, composed of nucleons and light clusters, to scrutinize how the presence of light clusters, emerging from few-body correlations, can affect the mean-field evolution and the development of spinodal instabilities (and associated growth times), eventually leading to the full disassembly of the system into pieces of various sizes.  

\paragraph{Theoretical framework.} 
Let us consider a system constituted by nucleons, namely neutrons $n$ and protons $p$, and one light cluster species $d$ introduced as explicit degree of freedom, in thermodynamical equilibrium at a temperature $T$ (the discussion can be easily extended to the presence of 
several light cluster species).  
Let us denote by $\rho_{b} = \sum_{j} \rho_{j} A_{j}$ the total baryon density, as expressed in terms of the densities $\rho_{j}$ and mass numbers $A_{j}$ of the three constituents ($j=n, p, d$) considered. 
The phase-space distribution functions $f_{j}$  
are given by 
\begin{equation}
f_{j}\left( \epsilon_{j}\right) = 
\left[ \exp\left(\dfrac{\epsilon_{j} - \mu_{j}^\ast}{T} \right) - (-1)^{A_{j}} \right]^{-1} 
\end{equation}
where $\mu_{j}^{\ast}$ is the effective chemical potential and $\epsilon_{j} = \dfrac{p^{2}}{2m_{j}}$, with $m_{j}$ denoting the mass of the considered constituent. The latter is defined as $m_{j} = A_{j}m - B_{j}$, where $m = 939$ MeV\footnote{A unitary value is adopted for the speed of light.} is the bare nucleon mass and  $B_{j}$ is the binding energy in vacuum of the cluster, obviously vanishing for free nucleons. The number density $\rho_{j}$ of each species is defined as  
\begin{equation}
\rho_{j} = g_{j} \int_{\Lambda_{j}} \dfrac{d \mathbf{p}}{(2\pi\hbar)^{3}} f_{j} 
\label{eq:rhoi}
\end{equation}
where $g_{j}$ is the spin-degeneracy and an (infra-red) cut-off momentum $\Lambda_{j}$ (Mott momentum) is introduced (only for the clusters, i.e., $\Lambda_{j} = 0$ for $j=n,p$) to take into account that, owing to in-medium effects related to Pauli-blocking, nuclear clusters can only form if their center of mass momentum is larger than the Mott one~\cite{ropkePRC2015, RoepkePRC2020, WangPRC2023}. 
In the most general case, the cut-off  $\Lambda_{j}$ depends on the densities of the constituents involved and on temperature, thus embedding effects similar to a (momentum-dependent) binding energy shift within a quasi-particle picture~\cite{ropkeNPA2011, burEPJA2022}. 
Within the present formalism, we assume that when clusters survive, they  keep the vacuum mass, regardless of the density of the surrounding medium. 
A consistent treatment, incorporating the effects related to the mass shift and to the  momentum cut-off, goes indeed beyond the scope of the present work.
Moreover, it is worth noting that correlations in the continuum, which might become important with increasing density~\cite{ropkePRC2015, burEPJA2022}, are also neglected here.
Moreover, it is worth noting that correlations in the continuum, which might become important with increasing density~\cite{ropkePRC2015, burEPJA2022}, are also neglected here.
 
The thermodynamical properties of the system can be fully characterized by its thermodynamical potential. At finite temperature, one has to consider the free-energy density $\mathcal{F} = \mathcal{E} - T \mathcal{S}$, where $\mathcal{S}$ is the entropy density and $\mathcal{E}$ is the energy density $\mathcal{E} = \mathcal{K} + \mathcal{U}$, expressed as the sum of the kinetic ($\mathcal{K}$) and potential ($\mathcal{U}$) terms [see Eqs.~(S1)-(S4) in the Supplemental Material]. In the framework of the energy density functional (EDF) theory, the latter is obtained from a (density-dependent) effective interaction. In this work we concentrate on temperature values $T \gtrsim 5$ MeV, which are relevant for fragmentation processes in heavy-ion collisions at Fermi/intermediate energies and also for several astrophysical scenarios\footnote{These temperature values mainly lie beyond the critical one for the transition to the Bose-Einstein condensate phase for the light clusters~\cite{wuJLTP2017}. The emergence of such a phase is further suppressed when introducing a momentum cut-off.}. Then the functional $\mathcal{F}$ does not include any contribution from boson condensation. 

Our aim here is to account, within a unified theoretical framework, for the formation of heavy fragments driven by the volume instabilities and the presence of light clusters. For that purpose, we undertake a linear response analysis of the collisionless (Vlasov) limit of the Boltzmann equation~\cite{ChomazPR2004, ColonnaPRC2008, burrelloPRC2019}, considering the interplay between nucleonic and light-cluster degrees of freedom and including in-medium effects. Then, by applying a small amplitude perturbation $\delta f_{j}$ to the distribution functions, starting from the initial condition $f_{j}$, the linearized Vlasov equations take the form
\begin{equation}
\label{eq:vlasov_linear}
\partial_{t} \left ( \delta f_{j} \right) + \nabla_\mathbf{r}( \delta f_{j}) \cdot \nabla_\mathbf{p} \varepsilon_{j}- \nabla_\mathbf{p} f_{j} \cdot \nabla_\mathbf{r} (\delta \varepsilon_{j}) = 0,
\end{equation}
where $\varepsilon_{j}$ is the single-particle energy, which is defined as the functional derivative of $\mathcal{E}$ 
\begin{equation}
\varepsilon_{j} \equiv \dfrac{(2\pi\hbar)^{3}}{g_{j}}\dfrac{\delta \mathcal{E}}{\delta f_{j} (\mathbf{p})}. 
\label{eq:varepsilon}
\end{equation}
One should note that, even in the  case of a density-dependent momentum cut-off that we will adopt hereafter, the mass-energy conservation laws are fully satisfied within the first-order approximation here considered. 
As shown in the Supplemental Material, the single-particle energy writes as $\varepsilon_{j} = \epsilon_{j} + U_{j} + \tilde{\varepsilon}_{j}^{\lambda}$, where $U_{j} = \dfrac{\partial \mathcal{U}}{\partial \rho_{j}}$ denotes the (momentum-independent) mean-field potential and the effective potential term
\begin{equation}
\tilde{\varepsilon}_{j}^{\lambda} = - \dfrac{\lambda_{d} + U_{d}}{1 + \Phi_{\lambda}^{dd} } \Phi_{\lambda}^{dj},\,{\rm with}\,\,\Phi_{\lambda}^{dj} = \alpha_{d}\sqrt{\lambda_{d}} f_{d}\left(\lambda_{d} \right) \dfrac{\partial \lambda_{d}}{\partial \rho_{j}},
\label{eq:Phi}
\end{equation}
with $\alpha_{d} = g_{d} \dfrac{\left(2 m_{d}\right)^{3/2}} {4 \pi^{2} \hbar^{3}}$, stems from the density dependence of the (cluster) kinetic energy cut-off $\lambda_{d}  = \dfrac{\Lambda_{d}^2}{2m_{d}}$. It is interesting to note that, by virtue of Eq.~\eqref{eq:rhoi}, one can write
\begin{equation}
\delta\rho_j(\mathbf{r},t) =  g_j\int_{\Lambda_{j}}\frac{d\mathbf{p}}{(2\pi\hbar)^3}\delta f_{j}-\delta_{jd} \sum_{l} \Phi_{\lambda}^{dl} \delta \rho_{l} 
\label{eq:deltarhojrt}
\end{equation}
where $\delta_{jd}$ denotes the Kronecker function.  The second term in the r.h.s. of Eq.~\eqref{eq:deltarhojrt} represents the change of light-cluster density due to the variations of the Mott momentum, triggered by density fluctuations. This change leads to the local appearance (or dissolution) of the light clusters, as their density adapts instantly to the local Mott momentum. From a dynamical point of view, this would imply a scenario in which the cluster formation (or dissolution) rate driven by the in-medium effects, $R_{d}$, is much larger than the changing rate, $R_{\delta\rho_{b}}$, of the local baryon density $\left( R_{d} \gg R_{\delta\rho_{b}}\right)$. 
However, to better highlight the impact of in-medium effects on the dynamics, we also consider the opposite case ($R_{d} \ll R_{\delta\rho_{b}}$), where the latter are neglected, assuming that the cut-off momentum remains constant during the propagation of density fluctuations ($\Phi_{\lambda}^{dj} = 0$). 

Equation~\eqref{eq:vlasov_linear} admits plane-wave solutions $\delta f_{j}$, periodic in time with frequency $\omega$ and wave vector $\mathbf{k}$, such as $
\delta f_{j} \sim \sum_{\mathbf {k}} \delta f_{j}^{\,\mathbf{k} }\, e^{ i (\mathbf{k}\cdot \mathbf{r} - \omega t)}$. Following a standard Landau procedure~\cite{landau1959theory}, one can derive a system of three coupled equations for neutron, proton and the light cluster species considered, which can be expressed in the following compact form (see Supplemental Material)
\begin{equation}
\delta \rho_{j} = 
- \chi_{j} \sum_{l} \left( F_{0}^{jl} + \tilde{F}_{\lambda}^{jl} \right) \delta \rho_{l} - \delta_{jd} \sum_{l} \Phi_{\lambda}^{dl} \delta \rho_{l},
\label{eq:deltarhoj}
\end{equation}
where $\chi_{j}$ denotes the Lindhard function, incorporating the momentum cut-off.  In Eq.\ \eqref{eq:deltarhoj}, we have introduced the parameters
\begin{equation}
F_{0}^{jl} = N_{j} \dfrac{\partial U_{j}}{\partial \rho_{l}}, \qquad \tilde{F}_{\lambda}^{jl} = N_{j} \dfrac{\partial \tilde{\varepsilon}_{j}^{\lambda}}{\partial \rho_{l}} \qquad j,l = n, p, d
\label{eq:landau}
\end{equation}
where the $\tilde{F}_{\lambda}^{jl}$ terms are defined in analogy with the standard Landau parameters $F_{0}^{jl}$, and $N_{j}$ indicates the thermally averaged level density (with the infra-red momentum cut-off). In the following, we will consider the simplest case of symmetric nuclear matter (SNM) with only deuterons added as explicit degrees of freedom. Indeed, although at low temperatures both microscopic quantum statistical (QS) and relativistic mean-field (RMF) calculations predict a clear dominance of $\alpha$ particles~\cite{typelPRC2010}, for SNM at $T \gtrsim 5$ MeV (the temperature regime of our interest) the leading role is played, to a large extent, by two-body correlations~\cite{wuJLTP2017, burEPJA2022}. 

\paragraph{Results.}
We adopt a simplified Skyrme-like effective interaction 
as in Ref.~\cite{baranPREP2005} and we assume that nucleons bound in deuterons feel the same mean-field potential as free nucleons. Other scenarios, proposed by recent works~\cite{qinPRL2012, paisPRC2019, burEPJA2022}, will be explored in an extended forthcoming paper. 

To parametrize the in-medium effects, we refer to the microscopic calculations of Ref.~\cite{ropkePRC2015} and we consider the following form for the kinetic energy cut-off
\begin{equation}
\lambda_{d} (\rho_{b}, T) = \beta_{d}\, \rho_{b}^{\gamma_{d}}\, \left [ 1 + \tanh \left( 1 - \xi_{d} \dfrac{\rho_{d}^{\rm Mott} (T)}{\rho_{b}} \right) \right]
\label{eq:EMt}
\end{equation}
with $\beta_{d} = 440$ MeV fm$^{2}$, $\xi_{d} = 2$ and $\gamma_{d}=2/3$, thus assuming a power-law dependence on the total baryon density~\cite{KuhrtsPRC2001, ropkeNPA2011, WangPRC2023}, properly smoothed around the Mott density $\rho_{d}^{\rm Mott}$ to avoid the emergence of discontinuities in the density derivatives of the cut-off. The adopted parameterization embeds moreover the temperature dependence of $\rho_{d}^{\rm Mott}$, as given in Ref.~\cite{ropkeNPA2011}. 

\begin{figure}[htp]
\centering
\includegraphics[width=8.5cm]{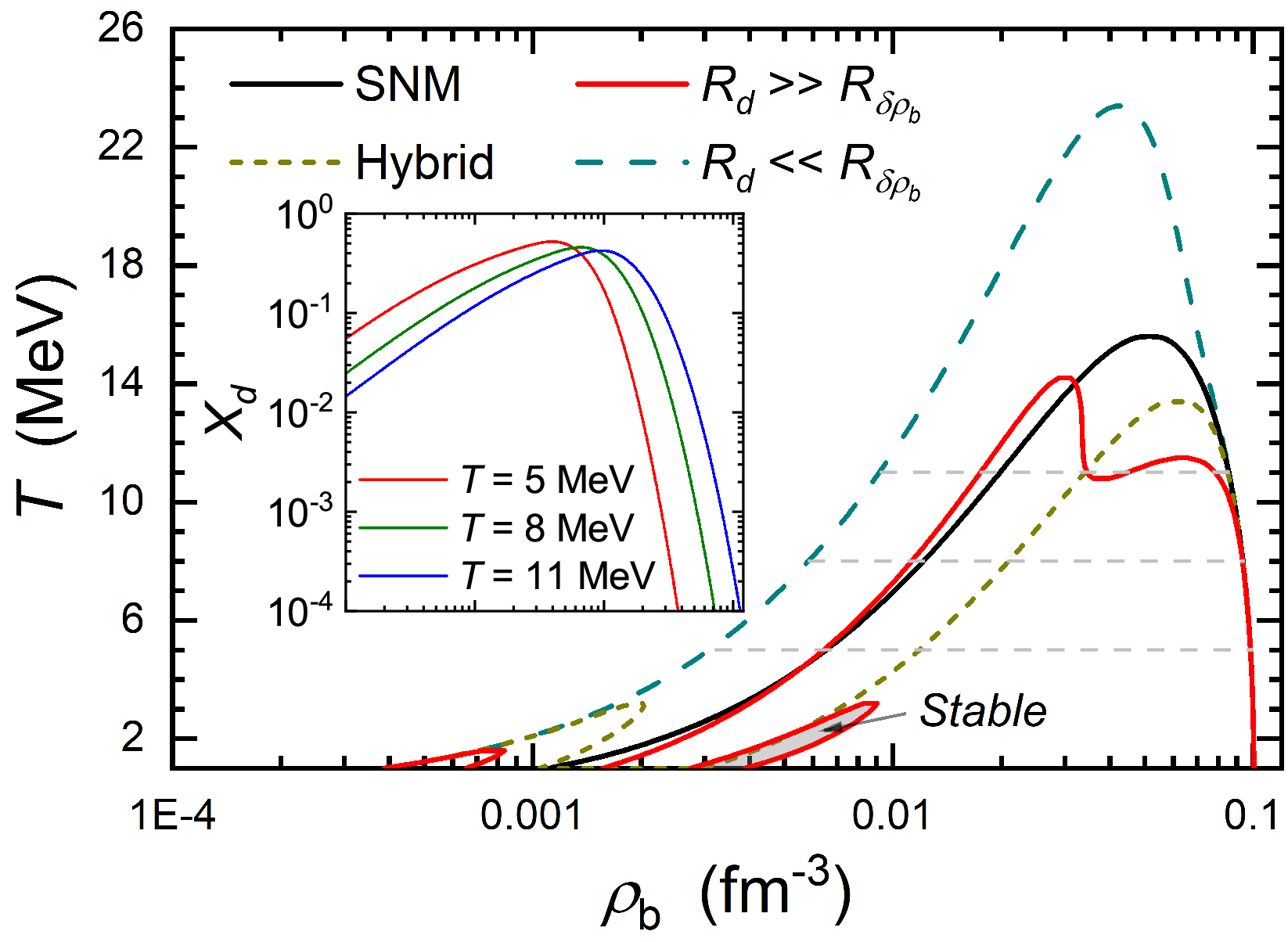}
\caption{Spinodal border in the $\left(\rho_{b}, T\right)$ plane in three cases: $1$) pure nucleonic matter (SNM, black); $2$) for nuclear matter with deuterons, including in-medium effects ($R_{d}\gg R_{\delta\rho_{b}}$, red); $3$) for nuclear matter with deuterons, neglecting ($R_{d}\ll R_{\delta\rho_{b}}$, cyan) in-medium effects in the dynamics. The result from a ``hybrid'' case is also included (green, see text for details). The inset shows the $\rho_{b}$ dependence of the deuteron fraction $X_{d}$ for three temperature values.}
\label{fig:border}
\end{figure}

As a reference point for our initial conditions, we impose chemical equilibrium. However, the latter condition is not necessarily reached during the expansion phase of a nuclear reaction and can be easily released in the calculations, without changing our conclusions. For the mean-field interaction and the density dependence of the cut-off here considered, the chemical equilibrium condition reduces to $\mu_{d}^{\ast} = \mu_{n}^{\ast} + \mu_{p}^{\ast} + B_{d}$, which allows one to fix the deuteron mass fraction $X_{d} = A_{d} \rho_{d}/\rho_{b}$ at each density. Such a quantity is shown in the inset of Fig.~\ref{fig:border}, as a function of the
total baryon density, for three temperature values. 
One may note that the plotted curves well reproduce the global trend predicted by QS or RMF calculations in Ref.~\cite{typelPRC2010}, as well as the cluster dissolution at increasing densities, thus validating the adopted choice for the ($\rho_{b}, T$) dependence of the cut-off. 

The onset of spinodal instabilities is identified by imposing that the determinant of the matrix associated with Eq.~\eqref{eq:deltarhoj} vanishes for  $\omega = 0$ ($\chi_{q} = \chi_{d} = 1)$~\cite{burPRC2014, ChomazPR2004}. The main panel of Fig.~\ref{fig:border} displays the spinodal border in the $(\rho_{b}, T)$ plane, as obtained by taking into account the local density-dependence of the cut-off in the dynamics (red line). This represents a suitable choice, considering that at the spinodal boundary one has $\omega=0$, therefore the condition $R_{d}\gg R_{\delta\rho_{b}}$ always holds~(note that for plane-wave solutions, one has $R_{\delta\rho_{b}}$ $\sim$ $-i\omega\delta\rho_{b}$). The curve is compared with the (cyan) one deduced by neglecting the density dependence of the cut-off ($\Phi_{\lambda}^{dj} = \tilde{F}_{\lambda}^{jl} = 0$), where  the relation defining the spinodal border is expressed as
\begin{equation}
\left(1 + F_{0} \right)\left( 1 + F_{0}^{dd} \right) - 2 F_{0}^{qd} F_{0}^{dq} = 0,    
\end{equation}
with $q=n$ or $p$ and $F_{0} = F_{0}^{nn} + F_{0}^{np}$, which leads to the usual pure nucleonic matter relation $(1 + F_{0})  = 0$ (black) for $N_{d}\to 0$, that is in absence of light clusters. For the sake of illustration, only in Fig.~\ref{fig:border}, we also plot (green) the curve related to a ``hybrid'' situation in which the density dependence of the cut-off is only neglected in the single-particle energies, thus imposing $\tilde{F}_{\lambda}^{jl} = 0$, while keeping $\Phi_{\lambda}^{dj} \ne 0$ in Eq.~\eqref{eq:deltarhoj}. It appears that including light clusters as explicit degrees of freedom has a strong effect on the extension of the spinodal region. If in-medium effects were neglected in the dynamics ($\Phi_{\lambda}^{dj}= 0)$, instabilities would occur over a wider region of the phase diagram, as a result of the stronger attraction generated by the deuteron mean-field potential, which enters the $F_{0}^{dd}$ contribution. On the other hand, a remarkable shrinkage of the unstable region is predicted in the hybrid case, because in-medium effects tend to increase the deuteron kinetic energy. Quite intriguingly, once in-medium effects are fully taken into account, i.e. including the $\tilde{F}_{\lambda}^{jl}$ terms (red), the spinodal border of the composite system remains closer to the one obtained for pure nucleonic matter (black).   
It is also interesting to observe that the delicate interplay between the deuteron attractive potential and in-medium effects is responsible for the emergence of small disjointed regions of instability at low temperature, both in the hybrid and full cases (see the region  below $0.002$ fm$^{-3}$). Moreover, in the case of the full calculations, a further escape and re-entrance from the region of spinodal instabilities is seen at a larger density, in 
some analogy with the findings of recent works~\cite{ROPKE2018224, VoskresenskyPPNP2023}, where the emergence of such a meta-stable region was discussed.

\begin{figure}[htp]
\centering
\includegraphics[width=8.5cm]{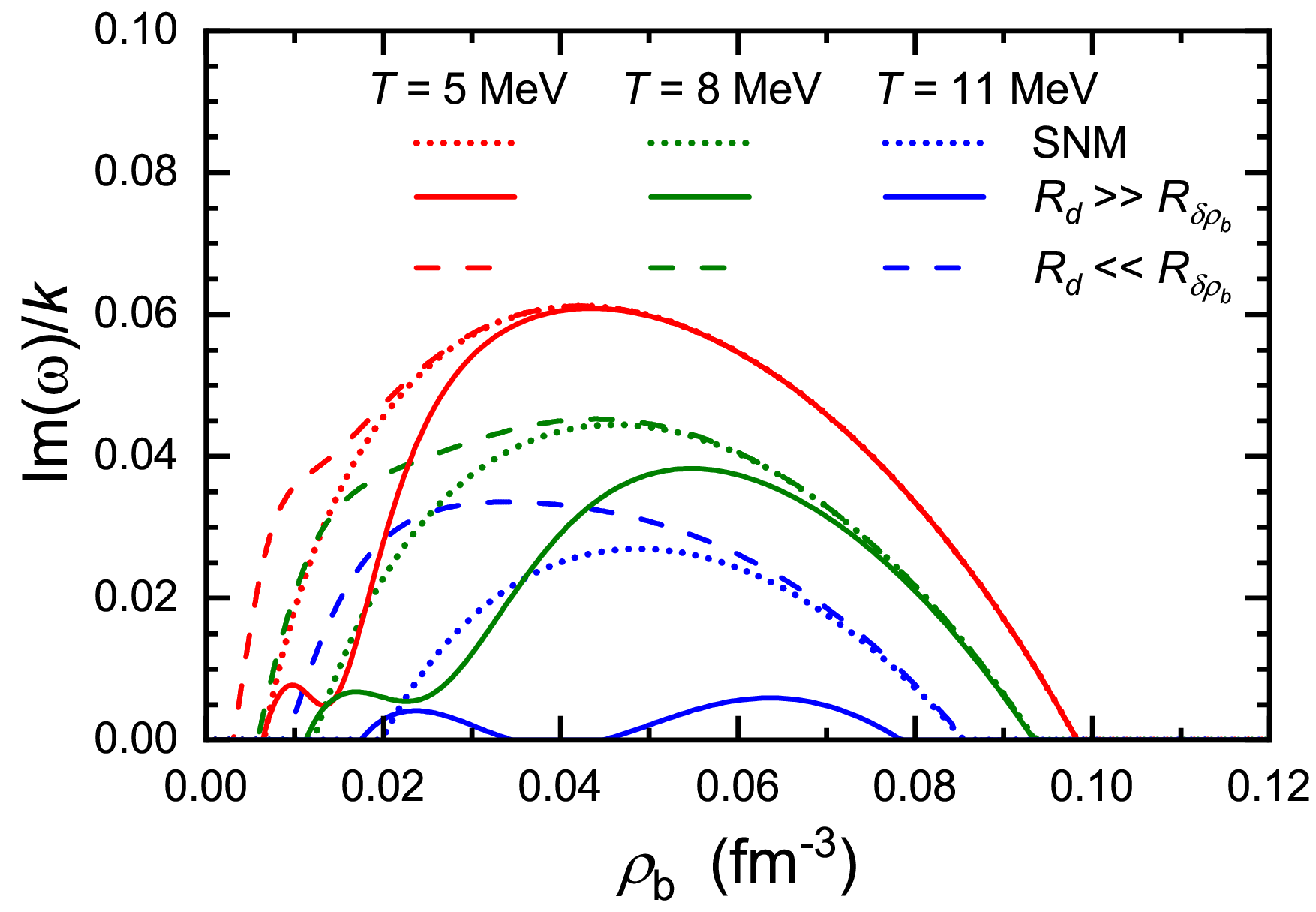}
\caption{Growth rate of the imaginary sound velocity, $\operatorname{Im} (\omega)/k$ (for $k \rightarrow 0$), as a function of the total baryon density $\rho_{b}$ for the same cases considered in Fig.~\ref{fig:border} and for three temperature values.}
\label{F:ImwIk_rh}
\end{figure}
In general, non-trivial solutions of $\delta\rho_l$ are obtained by requiring the determinant of the 
system associated with Eq.~\eqref{eq:deltarhoj} to be equal to zero. From this condition, one extracts the dispersion relation, connecting the frequency $\omega$ to the wave number ${k}$. Inside the spinodal region, a pure imaginary $\omega$ will be obtained in a certain $k$ interval, leading to an unstable growth of those modes. In Fig.~\ref{F:ImwIk_rh}, we show the imaginary sound velocity, $\operatorname{Im} (\omega)/k$ (for $k \rightarrow 0$), as a function of the total baryon density, for the same temperature values as in the inset of Fig.~\ref{fig:border}. We notice that in-medium effects associated with the density dependence of the cut-off along the dynamics {~($R_{d}\gg R_{\delta\rho_{b}}$)} typically induce a suppression of the growth rate of the instability with respect to the case of pure nucleonic matter, thus slowing down fragment formation, especially at the largest temperature considered, where deuterons survive up to larger densities. This suppression contrasts with the increase that would occur if the variation of the cut-off had been ignored during the fragmentation process~$\left(R_{d} \ll R_{\delta\rho_{b}}\right)$. Moreover, the full calculations lead to a non-trivial convex region in the density behavior of $\rm{Im}(\omega)/k$, which is connected to the features observed in  Fig.~\ref{fig:border} for the spinodal border (see in particular the results at $T$ around $11~\rm MeV$).

\begin{figure}[htp]
\centering
\includegraphics[width=8.5cm]{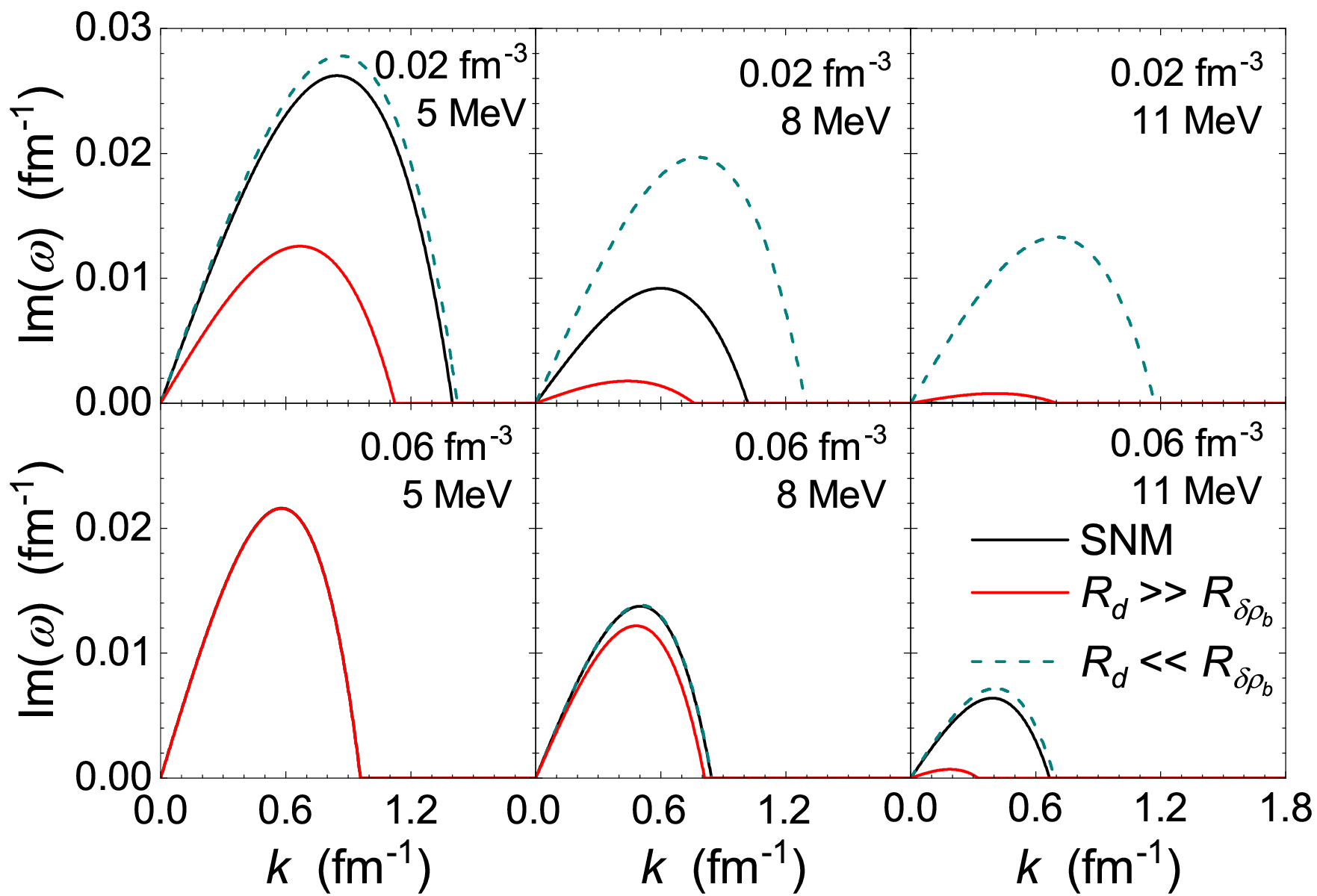}
\caption{Growth rate of the instability, $\rm{Im}(\omega)$, as a function of the wave number $k$, for the same cases as in Figs.~\ref{fig:border} and ~\ref{F:ImwIk_rh}, at various density and temperature values.}
\label{F:Imw_k}
\end{figure}

The dependence of the growth rate on the wave number, for two typical density values lying inside the spinodal region ($\rho_{b} = 0.02$ fm$^{-3}$ and $\rho_{b} = 0.06$ fm$^{-3}$) is displayed in Fig.~\ref{F:Imw_k}, for the same temperature values considered above. One notices that, due to the $k$-dependence of the Landau parameters, which arises from surface terms included in $\mathcal{U}$, the growth rate exhibits a maximum, which means that the system favors the growth of the density fluctuations with a given $k$. This feature has been used as an evidence to identify the spinodal decomposition in heavy-ion collisions~\cite{BorPLB2018}. As a general result, one may argue that taking into account the density-dependence of the in-medium effects~(corresponding to $R_{d}\gg R_{\delta\rho_{b}}$) has a strong impact on the maximum growth rate, which is quenched and shifted to lower $k$ values. This in turn induces an increase of  the average size of the fragments produced through the spinodal mechanism, especially when lower densities or higher temperature values are considered. Once again, the opposite scenario occurs when in-medium effects in the dynamics are neglected $\left(R_{d} \ll R_{\delta\rho_{b}}\right)$, since in this case the presence of light clusters leads to higher growth rates. For instance, for $\rho_b = 0.02$~fm$^{-3}$ and T = 8 MeV, the wavelength of the most unstable modes can deviate by $\approx \pm 20\%$~(with the sign depending on whether in medium-effects are considered or not) from the typical wavelength values ($\lambda\approx 10$~fm) observed in pure nucleonic matter~\cite{ChomazPR2004}. It is also interesting to note that light clusters do not affect much the growth of spinodal instabilities occurring at larger densities ($\rho_b \geqslant \rho_0/3$) and moderate temperatures ($T \leqslant 8$~MeV).

\begin{figure}[htp]
\centering
\includegraphics[width=8.5cm]{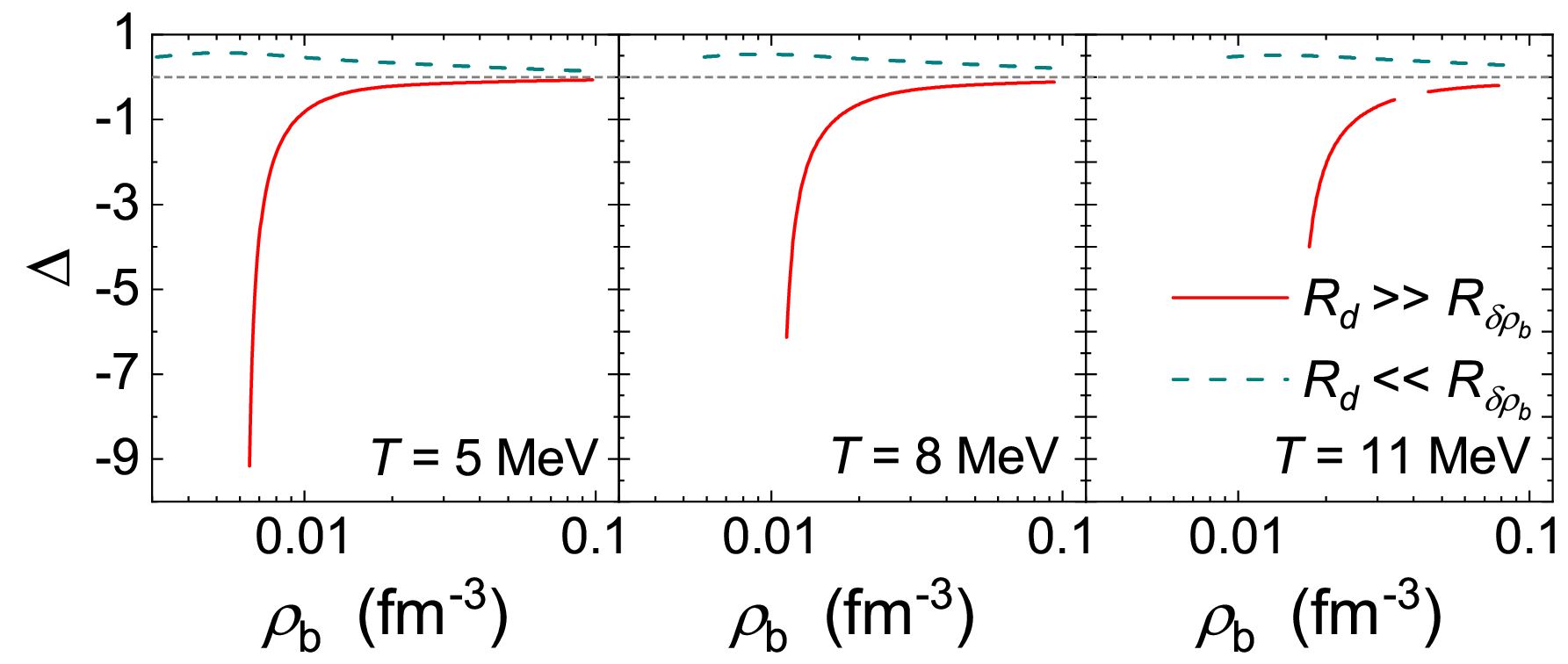}
\caption{The quantity $\Delta$ (see text) as a function of the total baryon density $\rho_{b}$ for nuclear matter with deuterons, neglecting (cyan) or including (red) in-medium effects in the dynamics, for three temperature values. Lines are drawn only inside the spinodal region.}
\label{F:mode}
\end{figure}
An in-depth insight into the direction of the unstable modes in the space of density fluctuations is provided by the ($\delta\rho_{S}/\delta\rho_{d})$ ratio, where $\rho_S = \rho_n + \rho_p$.
Figure~\ref{F:mode} exhibits the quantity $\Delta = (\delta\rho_{S}/\delta\rho_{d})/(\rho_{S}/\rho_{d})$ as a function of $\rho_{b}$ inside the spinodal region for the two options considered above (i.e., either neglecting or fully considering the density dependence of the cut-off). 
Positive (negative) values indicate that $\rho_{S}$ and $\rho_d$ fluctuations move in (out of) phase, respectively. One observes that, when assuming $R_{d} \ll R_{\delta\rho_{b}}$ (i.e., neglecting in-medium effects along the dynamics), light clusters move in phase with the nucleons, favoring the growth of instabilities and possibly 
contributing to the formation of massive fragments. Quite interestingly, once in-medium effects are taken into account $\left( R_{d}\gg R_{\delta\rho_{b}} \right)$, deuterons move out-of phase with respect to nucleons, thus migrating towards lower density regions while the nucleon density fluctuations grow and fragments emerge.   

\paragraph{Summary and outlook.}
Within a linearized Vlasov approach, we have explored the 
occurrence of spinodal instabilities in dilute nuclear matter with light-cluster degrees of freedom. We show that the presence of light clusters and, in particular, in-medium (Mott) effects on their propagation  have a crucial impact on the features of the unstable modes responsible for the system disassembly. Whereas, if in-medium effects in the dynamics are neglected, the light clusters move in phase with nucleons, cooperating in the formation of fragments, local in-medium effects induce a sort of distillation mechanism~\cite{baranPREP2005, burPRC2014}, with clusters moving towards the lower density regions, slowing down the instability growth and eventually leading to the dominance of different fragmentation modes. These findings underline the relevance of a proper inclusion of light-cluster degrees of freedom and of in-medium effects in the description of dilute nuclear systems, revealing important consequences for the understanding of heavy-ion collisions and astrophysical phenomena taking place in low-density and moderate-temperature environments~\cite{HuthNAT2022, tsangNAT2024}. 

\begin{acknowledgements}
\paragraph{Acknowledgements.}
Stimulating discussions with Gerd R\"{o}pke and Stefan Typel are gratefully acknowledged. The authors also thank Edoardo Lanza for valuable comments on this work.
\end{acknowledgements}

\newpage

\begin{widetext}
\begin{center}
    \large
    ~~\\
    \bf{Supplemental Material}
\end{center}

\section{Energy density functional and single particle energy}
\label{SM:EDF}
\noindent In the expression of the free energy density functional $\mathcal{F} = \mathcal{E} - T \mathcal{S}$, the energy density functional  $\mathcal{E} = \mathcal{K} + \mathcal{U}$  is the sum of the kinetic energy density
\begin{equation}
\mathcal{K}  =  
\sum_{j} g_{j} \int_{\Lambda_{j}} \dfrac{d \mathbf{p}}{(2\pi\hbar)^{3}} 
f_{j} \epsilon_{j} 
\end{equation}
and the potential energy density $\mathcal{U}$. 
For the latter, we adopt a (momentum-independent) Skyrme-like effective interaction, associated with an isoscalar potential energy density: 
\begin{equation}
\mathcal{U} = \frac{A}{2}\frac{\rho_{b}^2}{\rho_0} + \frac{B}{\alpha+2}\frac{\rho_{b}^{\alpha+2}}{\rho_0^{\alpha+1}} + \frac{D}{2}(\nabla_{\mathbf{r}}\rho_{b})^2, 
\end{equation}
where $A = -356$ MeV, $B = 303$ MeV, $D = 130$ MeV~fm$^{5}$  and $\alpha = 1/6$ are combinations of the standard Skyrme parameters, and $\rho_{0} = 0.16$ fm$^{-3}$ is the saturation density.
$\mathcal{S}$ denotes the entropy density, given by
\begin{equation}
\mathcal{S} = - \sum_{j} g_{j} \int_{\Lambda_{j}} \dfrac{d \mathbf{p}}{(2\pi\hbar)^{3}} \left[ f_{j} \ln f_{j} + \dfrac{1 - \sigma_{j} f_{j}}{\sigma_{j}} \ln \left ( 1 - \sigma_{j} f_{j} \right) \right]
\end{equation}
where $\sigma_{j} =  \pm 1$ for fermions and bosons, respectively. Then the free-energy density functional can be expressed as
\begin{equation}
\mathcal{F} =
\mathcal{U} + \sum_{j} \mu_{j}^{\ast} \rho_{j} -  T  
\sum_{j} \dfrac{g_{j}}{\sigma_{i }} \int_{\Lambda_{j}} \dfrac{d \mathbf{p}}{(2\pi\hbar)^{3}} \ln \left[ 1 + \sigma_{j} \exp\left( - \dfrac{\epsilon_{j} - \mu_{j}^{\ast}}{T}\right) \right],
\label{eq:functional}
\end{equation}
from which the chemical potentials $\mu_{j} = \dfrac{\partial \mathcal{F}}{\partial \mathcal \rho_{j}}$ can be derived. 

The variation of the energy density $\delta \mathcal{E}$ in the space of distribution function fluctuations $\delta f_{j}$
\begin{equation}
\delta \mathcal{E} \left [ \delta f_{n}, \delta f_{p}, \delta f_{d} \right ] = \delta \mathcal{K}\left [ \delta f_{n}, \delta f_{p}, \delta f_{d} \right ] + \delta \mathcal{U} \left [ \delta f_{n}, \delta f_{p}, \delta f_{d} \right ],
\end{equation}
is given in terms of the fluctuation $\delta \mathcal{K}$ of the kinetic energy density
\begin{equation}
\delta \mathcal{K} = \sum_{j} g_{j} \int_{\Lambda_{j}} \dfrac{d \mathbf{p}}{(2\pi\hbar)^{3}}  \epsilon_{j}\delta f_{j} - \lambda_{d}  \sum_{j} \Phi_{\lambda}^{dj} \delta \rho_{j}
\end{equation}
and of the fluctuation $\delta \mathcal{U}$ of the potential energy density
\begin{equation}
\delta \mathcal{U} =  
\sum_{j}\dfrac{\partial \mathcal{U}}{\partial \rho_{j}} \delta \rho_{j} =  \sum_{j} U_{j} \delta \rho_{j}
\end{equation}
where 
\begin{equation}
\Phi_{\lambda}^{dj} = \alpha_{d}\sqrt{\lambda_{d}} f_{d}\left(\lambda_{d} \right) \dfrac{\partial \lambda_{d}}{\partial \rho_{j}}
\end{equation}
with $\alpha_{d} = g_{d} \dfrac{\left(2 m_{d}\right)^{3/2}} {4 \pi^{2} \hbar^{3}}$ and $f_{d}\left(\lambda_{d} \right)$ denoting the cluster distribution function at equilibrium, 
evaluated at $\lambda_{d}  = \dfrac{\Lambda_{d}^2}{2m_{d}}$.
By taking into account that, by virtue of Eq.~(6) in the main text,
\begin{eqnarray}
\delta \rho_{j} \left( 1 + \delta_{jd} \Phi_{\lambda}^{dj} \right) & = &  g_j\int_{\Lambda_{j}}\frac{d\mathbf{p}}{(2\pi\hbar)^3}\delta f_{j} -\delta_{jd} \sum_{q=n,p} \Phi_{\lambda}^{dq} \delta \rho_{q} \nonumber \\
& = & g_j\int_{\Lambda_{j}}\frac{d\mathbf{p}}{(2\pi\hbar)^3}\delta f_{j} -
\delta_{jd} \sum_{q=n,p} \Phi_{\lambda}^{dq} g_{q}\int \frac{d\mathbf{p}}{(2\pi\hbar)^3}\delta f_{q},\end{eqnarray}
one finally obtains 
\begin{eqnarray}
\delta \mathcal{E}\left [ \delta f_{n}, \delta f_{p}, \delta f_{d} \right ] & = & \sum_{j} g_j\int_{\Lambda_{j}}\frac{d\mathbf{p}}{(2\pi\hbar)^3} \left( \epsilon_{j} + \dfrac{U_{j} - \lambda_{d} \Phi_{\lambda}^{dj}}{1 + \delta_{jd} \Phi_{\lambda}^{dj}}  \right)\delta f_{j} - \dfrac{U_{d} - \lambda_{d} \Phi_{\lambda}^{dd}}{1 +\Phi_{\lambda}^{dd}}\sum_{q=n,p} \Phi_{\lambda}^{dq} g_{q}\int \frac{d\mathbf{p}}{(2\pi\hbar)^3}\delta f_{q}, \nonumber \\
\end{eqnarray}
Then, the single-particle energy, as defined in Eq.~(4) of the main text, can be written as
\begin{equation}
\varepsilon_{j} = \epsilon_{j} + U_{j} + \tilde{\varepsilon}_{j}^{\lambda},
\end{equation}
where, beside the momentum-independent mean-field potential $U_{j}$, it appears the extra term
\begin{equation}
\tilde{\varepsilon}_{j}^{\lambda} = - \dfrac{\lambda_{d} + U_{d}}{1 + \Phi_{\lambda}^{dd} } \Phi_{\lambda}^{dj}
\end{equation}
playing the role of an effective potential contribution, which arises from the density dependence of the cluster kinetic energy cut-off $\lambda_{d}$. The variation of the single particle energy $\delta \varepsilon_{j}$ in the space of density fluctuations $\delta \rho_{j}$, entering the
linearized Vlasov equations (see Eq.~(3) in the main text), is then easily evaluated as: 
\begin{eqnarray}
\delta \varepsilon_{j} & = & \sum_{l} \dfrac{\partial U_{j}}{\partial \rho_{l}} \delta \rho_{l}  + \sum_{l} \dfrac{\partial \tilde{\varepsilon}_{j}^{\lambda}}{\partial \rho_{l}} \delta \rho_{l} \label{eq:deltavarepsilon}
\end{eqnarray}

\section{Linearized Vlasov equations and Landau parameters}
\label{SM:Landau}
\noindent Let us follow a standard Landau procedure, starting from the system of three coupled equations for the fluctuations of neutron, proton and light cluster distribution functions $\delta f_{j}$, given by Eq.~(3) in the main text. In the case of a momentum-independent interaction,
one has 
\begin{equation}
\nabla_{\mathbf{p}} \varepsilon_{j} = \dfrac{\mathbf{p}}{m_{j}} \qquad \nabla_{\mathbf{p}} f_{j} = \dfrac{\partial f_{j}}{ \partial \epsilon_{j}}\dfrac{\mathbf{p}}{m_{j}},
\end{equation}
and the linearized Vlasov equations reduce to:
\begin{equation}
\frac{\partial (\delta f_{j})}{\partial t} + \nabla_\mathbf{r}( \delta f_{j}) \cdot  \mathbf{v}_{j} - \dfrac{\partial f_{j}}{ \partial \epsilon_{j}} \mathbf{v}_{j} \cdot \nabla_\mathbf{r} (\delta \varepsilon_{j}) = 0,
\label{eq:vlasov_MI}
\end{equation}
where $\mathbf{v}_{j} = \dfrac{\mathbf{p}}{m_{j}}$ denotes the velocity. Then, for solutions of the type $\delta f_{j} \sim \sum_{\mathbf {k}} \delta f_{j}^{\,\mathbf{k} }\, e^{ i (\mathbf{k}\cdot \mathbf{r} - \omega t)}$ and exploiting Eq.~\eqref{eq:deltavarepsilon}, the system becomes
\begin{equation}
\left( - i \omega + i \mathbf{k} \cdot \mathbf{v}_{j} \right) \delta f_{j} - \dfrac{\partial f_{j}}{ \partial \epsilon_{j}} \mathbf{v}_{j} \cdot \nabla_\mathbf{r} \sum_{l} \left(\dfrac{\partial U_{j}}{\partial \rho_{l}} + \dfrac{\partial \tilde{\varepsilon}_{j}^{\lambda}}{\partial \rho_{l}} \right) \delta \rho_{l} = 0      
\end{equation}
and, taking into account the definition of the Landau parameters given by Eq.~(8) in the main text, one gets
\begin{equation}
\delta f_{j} = \dfrac{\partial f_{j}}{ \partial \epsilon_{j}} \dfrac{i \mathbf{k} \cdot \mathbf{v}_{j}}{\left( - i \omega + i \mathbf{k} \cdot \mathbf{v}_{j} - i0^{+} \right) } \dfrac{1}{N_{j}}\sum_{l} \left( F_{0}^{jl}+ \tilde{F}_{\lambda}^{jl} \right)\delta \rho_{l},    
\end{equation}
where
\begin{equation}
\label{eq:nj}
N_{j} = - g_{j} \int_{\Lambda_{j}}\frac{d\mathbf{p}}{(2\pi\hbar)^3}\frac{\partial f_{j}}{\partial\epsilon_{j}} 
\end{equation}
is the thermally averaged level density of the considered species $j$.

Finally, by taking the momentum integrals on both sides and bearing in mind the definition of the Lindhard function:
\begin{equation}
\chi_{j} \left (\omega, \mathbf{k} \right) = \dfrac{g_{j}}{N_{j}} \int_{\Lambda_{j}} \dfrac{ d\mathbf{p}}{\left( 2 \pi \hbar\right)^{3}} \dfrac{\mathbf{k} \cdot \mathbf{v}_{j}}{\omega - \mathbf{k} \cdot \mathbf{v}_{j} + i0^{+}} \dfrac{\partial f_{j}}{ \partial \epsilon_{j}}, 
\label{eq:lindhard}
\end{equation}
one may express the system in the following compact form  
\begin{equation}
\delta \rho_{j} = 
- \chi_{j} \sum_{l} \left( F_{0}^{jl} + \tilde{F}_{\lambda}^{jl}\right) \delta \rho_{l} - \delta_{jd} \sum_{l} \Phi_{\lambda}^{dl} \delta \rho_{l}, 
\label{eq:system}
\end{equation}
or, more expicitly, as:
\begin{alignat}{4}[left = \empheqlbrace]
\label{eq:system_npd}
\left[ 1 + \left(F_{0}^{nn} + \tilde{F}_{\lambda}^{nn} \right) \chi_{n} \right] \delta \rho_{n} & {}+{} & \left(F_{0}^{np} + \tilde{F}_{\lambda}^{np} \right)\chi_{n} \delta \rho_{p}   & {}+{} & \left(F_{0}^{nd} + \tilde{F}_{\lambda}^{nd} \right) \chi_{n} \delta \rho_{d} & {}={} & 0  \nonumber \\
\left(F_{0}^{pn} + \tilde{F}_{\lambda}^{pn} \right) \chi_{p} \delta \rho_{n} & {}+{} & \left[ 1 + \left(F_{0}^{pp} + \tilde{F}_{\lambda}^{pp} \right)\chi_{p} \right] \delta \rho_{p}  & {}+{} & \left(F_{0}^{pd} + \tilde{F}_{\lambda}^{pd} \right) \chi_{p} \delta \rho_{d}& {}={} & 0  \nonumber \\
\left[ \left(F_{0}^{dn} + \tilde{F}_{\lambda}^{dn} \right) \chi_{d} + \Phi_{\lambda}^{dn} \right] \delta \rho_{n} & {}+{} &  \left[ \left(F_{0}^{dp} + \tilde{F}_{\lambda}^{dp} \right) \chi_{d} + \Phi_{\lambda}^{dp} \right] \delta \rho_{p} & {}+{} & \left[ 1 + \left(F_{0}^{dd} + \tilde{F}_{\lambda}^{dd} \right) \chi_{d} + \Phi_{\lambda}^{dd} \right] \delta \rho_{d} & {}={} & 0 
\end{alignat}
In case of SNM with deuterons, by introducing $\delta \rho_{S} = \delta \rho_{n} + \delta \rho_{p}$, the system of Eq.~\eqref{eq:system_npd} finally reduces to
\begin{alignat}{4}[left = \empheqlbrace]
\left[ 1 + \left( F_{0} + 2\tilde{F}_{\lambda}^{qq} \right)\chi_{q} \right] \delta \rho_{S}  & {}+{} & 2 \left(F_{0}^{qd} + \tilde{F}_{\lambda}^{qd} \right) \chi_{q} \delta \rho_{d}  & {}={} & 0 \nonumber \\
\left[ \Phi^{dq}_{\lambda}  + \left( F_{0}^{dq} + \tilde{F}_{\lambda}^{dq} \right) \chi_{d} \right]  \delta \rho_{S} & {}+{} & \left[ 1 + \Phi^{dd}_{\lambda} + \left( F_{0}^{dd} + \tilde{F}_{\lambda}^{dd} \right) \chi_{d} \right] \delta \rho_{d} & {}={}&  0 
\end{alignat}
where $F_{0} = F_{0}^{qq} + F_{0}^{qq^{\prime}}$, $q, q^{\prime} = n, p$ with $q \ne q^{\prime}$.

\section{Parameterization for the deuteron Mott momentum}
\label{SM:parameterizations}

In Fig.~\ref{F:PMt}, we show the deuteron cut-off momentum $\Lambda_{d} = \sqrt{2 m_{d} \lambda_{d}}$ as a function of the cubic root of the total baryon density, $(\rho_{b}/\rho_0)^{1/3}$ with $\rho_0$ $=$ $0.16~\rm fm^{-3}$, as obtained by adopting the parameterization given in Eq.~(9) in the main text, for $T$ $=$ $5~\rm MeV$, $8~\rm MeV$, and $11~\rm MeV$, respectively.

\begin{figure}[htp]
\centering
\includegraphics[width=8.5cm]{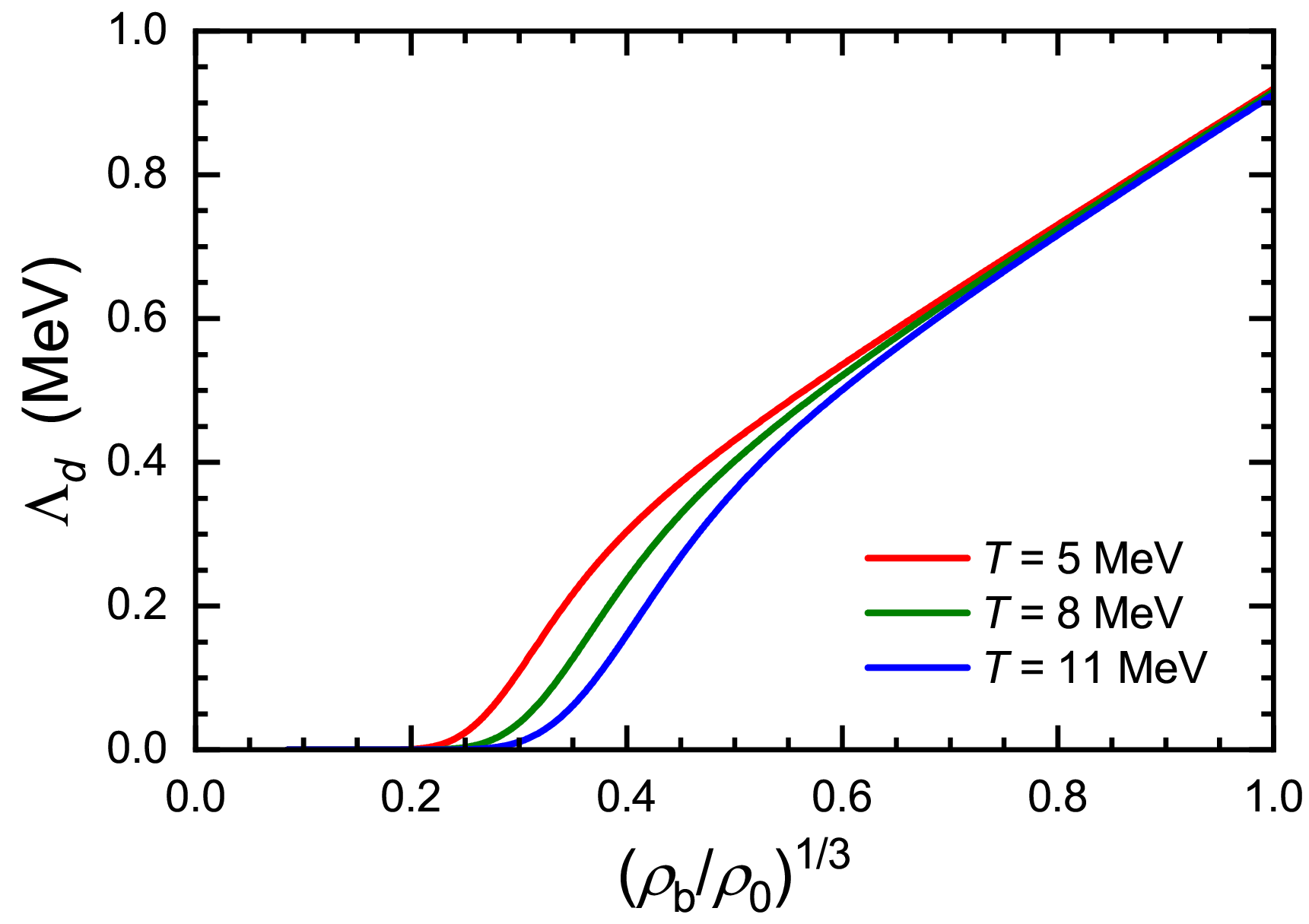}
\caption{Mott momentum of deuteron $\Lambda_d$ obtained by Eq.~(9) as a function of the cubic root of the total baryon density, $(\rho_{b}/\rho_0)^{1/3}$ with $\rho_0$ $=$ $0.16~\rm fm^{-3}$, 
for three temperature values. }
\label{F:PMt}
\end{figure}

\end{widetext}

\end{document}